\documentclass[superscriptaddress,amsmath,amssymb,aps,prb,twocolumn,nofootinbib,nolongbibliography]{revtex4-2}
\usepackage{graphicx}         
\usepackage{dcolumn}          
\usepackage{bm}               
\usepackage{upgreek}
\usepackage{booktabs} 				
\usepackage{tabularx}
\usepackage{array}
\usepackage[colorlinks=true,urlcolor=blue,breaklinks=true]{hyperref}

\usepackage{xcolor}

\usepackage{orcidlink}

\usepackage{xr}

\usepackage{siunitx}
\usepackage{stackengine}
\usepackage{svg}
\usepackage{graphicx,subfigure}

\definecolor{green2}{RGB}{15, 117, 19}

\begin{document}

\title{In-plane magnetic field driven conductance modulations in topological insulator kinks}

\author{Gerrit Behner\,\orcidlink{0000-0002-7218-3841}}
\email{g.behner@fz-juelich.de}
\affiliation{Peter Gr\"unberg Institut (PGI-9), Forschungszentrum J\"ulich, 52425 J\"ulich, Germany}
\affiliation{JARA-Fundamentals of Future Information Technology, J\"ulich-Aachen Research Alliance, Forschungszentrum J\"ulich and RWTH Aachen University, Germany}

\author{Kristof Moors\,\orcidlink{0000-0002-8682-5286}}
\email{k.moors@fz-juelich.de}
\affiliation{Peter Gr\"unberg Institut (PGI-9), Forschungszentrum J\"ulich, 52425 J\"ulich, Germany}
\affiliation{JARA-Fundamentals of Future Information Technology, J\"ulich-Aachen Research Alliance, Forschungszentrum J\"ulich and RWTH Aachen University, Germany}

\author{Yong Zhang}
\altaffiliation{On leave from: School of Information Science and Technology,
ShanghaiTech University, Shanghai, China}
\affiliation{Peter Gr\"unberg Institut (PGI-9), Forschungszentrum J\"ulich, 52425 J\"ulich, Germany}
\affiliation{JARA-Fundamentals of Future Information Technology, J\"ulich-Aachen Research Alliance, Forschungszentrum J\"ulich and RWTH Aachen University, Germany}

\author{Michael Schleenvoigt\,\orcidlink{0000-0002-2384-0366}}
\affiliation{Peter Gr\"unberg Institut (PGI-9), Forschungszentrum J\"ulich, 52425 J\"ulich, Germany}
\affiliation{JARA-Fundamentals of Future Information Technology, J\"ulich-Aachen Research Alliance, Forschungszentrum J\"ulich and RWTH Aachen University, Germany}

\author{Alina Rupp
\,\orcidlink{0009-0009-6140-4387}}
\affiliation{Peter Gr\"unberg Institut (PGI-9), Forschungszentrum J\"ulich, 52425 J\"ulich, Germany}
\affiliation{JARA-Fundamentals of Future Information Technology, J\"ulich-Aachen Research Alliance, Forschungszentrum J\"ulich and RWTH Aachen University, Germany}

\author{Erik Zimmermann
\,\orcidlink{0000-0002-1159-2027}}
\affiliation{Peter Gr\"unberg Institut (PGI-9), Forschungszentrum J\"ulich, 52425 J\"ulich, Germany}
\affiliation{JARA-Fundamentals of Future Information Technology, J\"ulich-Aachen Research Alliance, Forschungszentrum J\"ulich and RWTH Aachen University, Germany}

\author{Abdur Rehman Jalil
\,\orcidlink{0000-0003-1869-2466}}
\affiliation{Peter Gr\"unberg Institut (PGI-9), Forschungszentrum J\"ulich, 52425 J\"ulich, Germany}
\affiliation{JARA-Fundamentals of Future Information Technology, J\"ulich-Aachen Research Alliance, Forschungszentrum J\"ulich and RWTH Aachen University, Germany}

\author{Peter Schüffelgen
\,\orcidlink{0000-0001-7977-7848}}
\affiliation{Peter Gr\"unberg Institut (PGI-9), Forschungszentrum J\"ulich, 52425 J\"ulich, Germany}
\affiliation{JARA-Fundamentals of Future Information Technology, J\"ulich-Aachen Research Alliance, Forschungszentrum J\"ulich and RWTH Aachen University, Germany}

\author{Hans L\"uth}
\affiliation{Peter Gr\"unberg Institut (PGI-9), Forschungszentrum J\"ulich, 52425 J\"ulich, Germany}
\affiliation{JARA-Fundamentals of Future Information Technology, J\"ulich-Aachen Research Alliance, Forschungszentrum J\"ulich and RWTH Aachen University, Germany}

\author{Detlev Gr\"utzmacher}
\affiliation{Peter Gr\"unberg Institut (PGI-9), Forschungszentrum J\"ulich, 52425 J\"ulich, Germany}
\affiliation{JARA-Fundamentals of Future Information Technology, J\"ulich-Aachen Research Alliance, Forschungszentrum J\"ulich and RWTH Aachen University, Germany}

\author{Thomas Sch\"apers\,\orcidlink{0000-0001-7861-5003}}
\email{th.schaepers@fz-juelich.de}
\affiliation{Peter Gr\"unberg Institut (PGI-9), Forschungszentrum J\"ulich, 52425 J\"ulich, Germany}
\affiliation{JARA-Fundamentals of Future Information Technology, J\"ulich-Aachen Research Alliance, Forschungszentrum J\"ulich and RWTH Aachen University, Germany}

\hyphenation{}
\date{\today}

\begin{abstract}
We present low-temperature magnetoconductance measurements on Bi$_{1.5}$Sb$_{0.5}$Te$_{1.8}$Se$_{1.2}$ kinks with ribbon-shaped legs. The conductance displays a clear dependence on the in-plane magnetic field orientation. The conductance modulation is consistent with orbital effect-driven trapping of the topological surface states on different side facets of the legs of the kink, which affects their transmission across the kink. This magnetic field-driven trapping and conductance pattern can be explained with a semiclassical picture and is supported by quantum transport simulations. The interpretation is corroborated by varying the angle of the kink and analyzing the temperature dependence of the observed magnetoconductance pattern, indicating the importance of phase coherence along the cross section perimeter of the kink legs. 
\end{abstract}

\maketitle

\section{Introduction}

Three-dimensional topological insulators (3D TIs) are a class of materials in which strong spin-orbit coupling leads to a band inversion in the bulk electronic band structure \cite{Hasan10,Qi11}. This in turn causes the appearance of gapless surface states protected by time-reversal symmetry. Topological insulators are particularly interesting for applications in topological quantum computer architectures \cite{Nayak08,Alicea2012,Hyart13,Sarma15,Aasen16}. These circuits typically consist of networks of topological insulator nanoribbons combined with superconducting electrodes~\cite{Cook11, Cook2012, Manousakis2017, Juan2019, Legg2021, Legg2022, Heffels2023}. In this context, the electronic transport behavior of nanoribbons as well as more complex structures such as kinks and junctions of nanoribbons is of high interest.

In nanoribbon structures the existence of topological surface states is revealed by magnetotransport measurements \cite{Peng10,Xiu11,Dufouleur13,Arango16,Jauregui15,Koelzer20,Rosenbach20,Kim20}. Here, regular Aharonov-Bohm oscillations are observed when an axial magnetic field is applied, due to the presence of closed-loop surface states. The appearance of Aharonov-Bohm oscillations also confirms that the transport in the surface states is phase coherent with a phase-coherence length on the order of a few hundred nanometers~\cite{Dufouleur13,Arango16,Rosenbach20,Kim20}. For three-dimensional topological insulators, the surface transport is often accompanied by a transport channel carried by bulk carriers due to intrinsic doping effects \cite{Scalon12,Lostak89}. In this respect, nanostructures offer an unique advantage as their surface-to-volume ratio increases and the bulk and surface conductance contributions can be disentangled in transport experiments. In 3D TI nanoribbon-based three-terminal junctions it has been shown experimentally that the conductance can be steered by applying an in-plane magnetic field \cite{Koelzer21}. This steering effect is attributed to the interplay between the phase-coherent transport in topological surface states and an orbital effect on the side facets of the nanoribbon. This causes electrons in the surface states to be trapped on the upper or lower surface of a nanoribbon, depending on the relative orientation of each nanoribbon leg with respect to the in-plane magnetic field. 

Analogously to the effect of the in-plane field on the conductance in three-terminal junctions, the theoretical models also predict a $\pi$-periodic change in the conductance of a 3D TI nanoribbon-based kink under rotation of an in-plane magnetic field \cite{Moors18}. To address this issue we have studied the low-temperature magnetotransport properties of quadternary Bi$_{1.5}$Sb$_{0.5}$Te$_{1.8}$Se$_{1.2}$ kinks with different angles between the input and output terminal. Quadternary materials have been show to suppress the bulk conductivity in previous studies \cite{Kim2011,Zhi2011,Arakane2012}.The devices were fabricated using a selective-area molecular beam epitaxy (MBE) approach. The modulation of the conductance was then measured as a function of the angle of the applied in-plane magnetic field with respect to the orientation of the device. The experimental results are interpreted on the basis of a semi-classical theoretical model and corresponding simulations \cite{Koelzer21}. 

\section{Methods}

\begin{figure}[tb]
    \centering
    \includegraphics[width=0.99\linewidth]{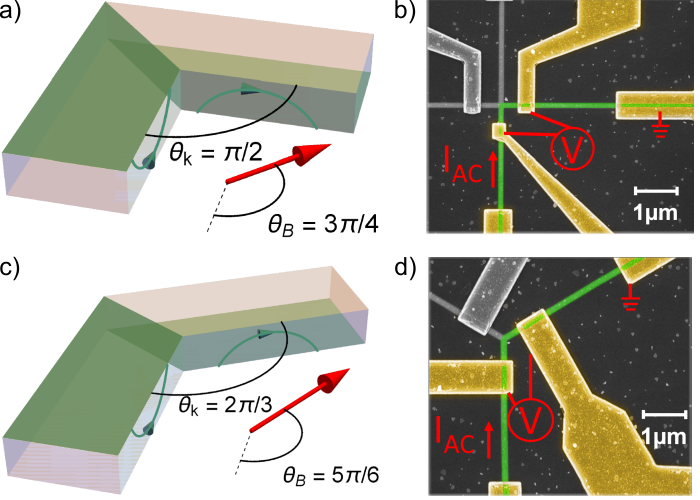}
    \caption{a) Schematic of the $\SI{90}{{}^\circ}$ kink. The red arrow indicates the orientation of the magnetic field with respect to the kink structure. The magnetic field angle $\theta_B$ here set to $\theta_B = 3 \pi/4$ which is the position for minimum transmission in the kink structure. The green surfaces on the legs of the kink indicate the trapping on opposite sides (reducing the transmission) of the surface-state channels due to the orbital effect of the external magnetic field. b) Scanning electron micrograph of the $\SI{90}{{}^\circ}$ kink devices. Note that the spare electrodes (in gray) were kept as open contacts in order to avoid any interference. The red annotations indicate the measurement setup. c) Corresponding schematic of the $\SI{120}{{}^\circ}$ kink with the red arrow indicating the magnetic field angle of $\theta_B = 5 \pi/6$. d) Scanning electron micrograph of the $\SI{120}{{}^\circ}$ kink devices.}
    \label{fig:Device}
\end{figure}

The Bi$_{1.5}$Sb$_{0.5}$Te$_{1.8}$Se$_{1.2}$ layer was grown by molecular beam epitaxy. A selective-area growth (SAE) approach was employed to yield the desired structures \cite{Kampmeier16,Koelzer20,Jalil23}. The samples were prepared in a 3 Step process. First, the substrate is prepared with a \SI{5}{nm}-thick-thermal SiO$_2$ layer and a \SI{20}{nm} thick PECVD Si$_3$N$_4$ layer to form the selective-area growth mask. The  shape of the SAE-trench was defined using reactive ion etching (CHF$_3$/O$_2$) and hydrofluoric acid wet etching after an electron beam lithography step. This is done in order to reveal the Si(111) surface as well as to passivate it before the growth. To prevent oxidation the film was capped using a 5-nm-thick AlO$_x$ layer. The kink structure are contacted with \SI{70}{nm} Ti contacts via ex-situ electron beam evaporation. Structures with a kink angle of \SI{90}{^\circ} and \SI{120}{^\circ} were investigated, which are part of a cross and symmetric 3-terminal structure, respectively. Figures~\ref{fig:Device} a) and c) show the corresponding schematic illustration of the kink structures together with the definitions of the kink $\theta_K$ and magnetic field angles $\theta_B$, respectively. In Figs.~\ref{fig:Device} b) and d) scanning electron beam micrographs of the \SI{90}{^\circ} and the  \SI{120}{^\circ} kink structure are shown.

From Hall measurements at 1.5 K we determined a carrier concentration of $2.3 \times 10^{13} \mathrm{cm}^{-2}$ and a mobility of $\SI{155}{cm^2/Vs}$ (see Supplemental Note 1 of Ref.~\cite{Zimmermann23}). These properties are attributed to a coexistence of diffusive bulk- and quasi-ballistic surface states in the material \cite{Behner23}. Transport in the micrometer sized Hall-devices is dominated by the diffusive bulk states.  The Ohmic contacts are formed by a $\SI{70}{nm}$ thick Ti layer. Before deposition of the metal layer the AlO$_x$ capping in the contact areas was removed by wet chemical etching and argon sputtering. 

The measurements were carried out in a variable temperature insert with a base temperature of \SI{1.5}{K}. The conductance of the device was measured using a standard four-probe lock-in setup (cf. Figs.~\ref{fig:Device}b) and d)). The rotation of the in-plane magnetic field with respect to the kink structure was realized by employing a mechanically rotatable sample rod, where the sample was rotated by $\pi$ in 19 steps. After each step the magnetic field was swept from \SI{-12}{T} to \SI{12}{T} in order to realize an effective rotation of the magnetic field of $2\pi$.

The simulations of the conductance in the kink structures are based on a semiclassical theoretical model explained in detail in Ref.~\cite{Koelzer21}. We consider a subband-quantized Dirac surface-state spectrum that is appropriate for the cross section geometry of the 3D TI kink legs and a Dirac point that is separated from the Fermi level by 0.1 eV.

\section{Magnetoconductance}

\begin{figure*}[htb]
    \centering
    \includegraphics[width=1\linewidth]{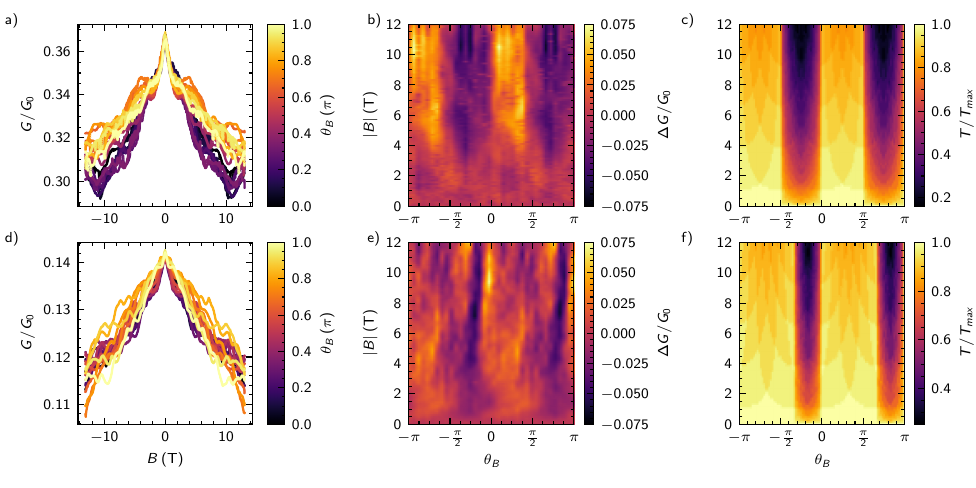}
    \caption{Conductance of the \SI{90}{^\circ} and \SI{120}{^\circ} kink structures at \SI{1.5}{K}. a) Conductance of the \SI{90}{^\circ} kink in units of $G_0=2e^2/h$ as a function of the in-plane magnetic field. The single lines are color coded where the color represents the angle under which the magnetic field is applied with respect to the device. b). Color map of the effective conductance $\Delta G/G_0 $ (see Eq.~\eqref{eq:effective-cond}) of the \SI{90}{^\circ} kink as a function of magnetic field angle $\theta_B$ and absolute magnetic field $|B|$. c) Calculated normalized transmission $T/T_\mathrm{max}$ as a function of absolute magnetic field $|B|$ and magnetic field angle $\theta_B$. d) to f) Corresponding plots for the \SI{120}{^\circ} kink.}
    \label{fig:BSTS-Kink-90DEG}
\end{figure*}

Figure~\ref{fig:BSTS-Kink-90DEG} a) shows the measured conductance in units of $G_0$ of the \SI{90}{^\circ} kink as a function of magnetic field, with $G_0=2e^2/h$ and $e$ the electron charge and $h$ Planck's constant. The measurements were taken at a temperature of \SI{1.5}{K}. Each color coded line represents a single measurement in between \SI{-12}{T} to \SI{12}{T} for a different in-plane magnetic field angle ranging from 0 to $\pi$. The definition of the in-plane magnetic field angle $\theta_B$ with respect to the sample is given in Fig.~\ref{fig:Device} a). 

All magnetoconductance traces shown in Fig.~\ref{fig:BSTS-Kink-90DEG} a) exhibit several prominent features. The conductance peak at zero magnetic field can be attributed to the weak antilocalization effect \cite{Hikami80}. This peak structure has previously been observed in topological insulator nanoribbon structures \cite{Koelzer20}. It is due to electron interference after scattering on impurities in combination with the strong spin-orbit coupling of the material. The small fluctuations of the conductance over larger magnetic field intervals represent universal conductance fluctuations \cite{Lee87}. These are caused by the interference of a finite number of trajectories due to the small dimensions of the sample. Apart from these two features observed in each curve the color coded magnetic field sweeps also reveal a change in resistance with a change in the magnetic field angle $\theta_B$. 
In order to analyzed the dependence of the magnetoconductance on the magnetic field orientation in detail, the data is plotted in the style of a color map as a function of the absolute magnetic field $|B|$ and the magnetic field angle $\theta_B$. Here, we consider only the effective conductance change $\Delta G/G_0 $ defined by
\begin{equation}
    \Delta G/G_0 = [G(B, \theta_B) - \langle G(B, \theta_B)\rangle_{\theta_B}]/G_0, 
    \label{eq:effective-cond}
\end{equation}
which is the normalized difference between the conductance at a certain magnetic field and in-plane field angle $G(B, \theta_B)$ and the average conductance at constant magnetic field averaged over all angles $\langle G(B, \theta_B) \rangle_{\theta_B}$. Figure~\ref{fig:BSTS-Kink-90DEG} b) shows the effective conductance for the \SI{90}{^\circ} kink. A clear $\pi$-periodic variation of the conductance with a variation of the in-plane magnetic field angle is visible. The areas of maximum and minimum conductance are centered around the values of  $-\pi/4$, $3\pi/4$ and $\pi/4$, $-3\pi/4$, respectively. The $\theta_B=3\pi/4$ field orientation for the minimum conductance case is indicated in Fig.~\ref{fig:Device} a). In order to demonstrate that the change in magnetoconductance is not a result of an angular dependence of universal conductance fluctuations (UCFs), a detailed analysis of the root-mean-square value of the conductance fluctuations $\delta_{G_\mathrm{RMS}}$ is provided in Supplemental Note 2 of the Supplemental Material. We refer the reader to Ref.~\cite{Koelzer21} for an analysis of the angle dependence of UCFs in a similar material system. \\  

In the following, a kink with a larger angle between in and output terminal, i.e., a  \SI{120}{^\circ} kink, is analysed. The corresponding normalized magnetoconductance as a function of magnetic field for in-plane angles between 0 and $\pi$ is shown in Fig.~\ref{fig:BSTS-Kink-90DEG} d). Similar to the conductance of the \SI{90}{^\circ} kink a clear modulation of the conductance with the variation of the in-plane field angle can be seen. Analogous to the first device this data is converted to the effective conductance change using Eq.\,(\ref{eq:effective-cond}). Figure~\ref{fig:BSTS-Kink-90DEG} e) shows the effective magnetoconductance of the \SI{120}{^\circ} kink. A shift of the position of the minimum and maximum conductance compared to measurements shown in Fig.~\ref{fig:BSTS-Kink-90DEG} b) for the \SI{90}{^\circ} kink is observed. The experimentally observed modulations match well with the theoretically expected positions for maximum and minimum conductance, i.e. $-\pi/6$, $5\pi/6$ and $-2\pi/3$, $\pi/3$, respectively. The positions are derived from trivial geometrical considerations regarding the aligned and transverse orientation of the magnetic field with respect to the kink-angle.

To explain the origin of the angle-modulated conductance pattern, we consider the impact of the orbital effect on the topological surface states in a semiclassical picture \cite{Koelzer21}, analogous to the treatment of the steering effect in a topological insulator-based T-junction in Ref.~\cite{Moors18}. On the top and bottom facets of the ribbon, the Lorentz force points perpendicular to the surface and can be neglected for surface-state motion that is bound to the surface of the ribbon. On the side facets (assumed to be perpendicular to the plane here), however, the Lorentz force induces a circular motion of the charge carriers (see Fig.~\ref{fig:Device} a) and c)). Depending on the transverse velocity of the surface state, the carriers will be able to traverse the side facet from bottom to top (or vice versa) or not. Thus, when the transverse extent is smaller than the height of the ribbon, the carriers cannot traverse the side facet. For surface states that wrap phase-coherently around the perimeter of the ribbon cross section, it follows that the top or bottom surface effectively gets  depleted, or, equivalently, carriers are trapped on the opposite surface. When the carriers on the two legs of the kink are trapped on opposite facets of the ribbon, the transmission across the kink is suppressed, as illustrated in Figs.~\ref{fig:Device} a) and c). This is also demonstrated by quantum transport simulations in the  Supplemental Material of Ref.~\cite{Koelzer21}. In Figs.~\ref{fig:BSTS-Kink-90DEG} c) and f) the calculated normalized transmission $T/T_\mathrm{max}$ as a function of magnetic field angle $\theta_B$ and absolute magnetic field $|B|$ are given for the \SI{90}{^\circ} kink and \SI{120}{^\circ} kink structures, respectively (see Supplemental Note 1 for details). By comparing to Figs.~\ref{fig:BSTS-Kink-90DEG} b) and e) one finds that the experimental results nicely follow the transmission pattern that results from the semiclassical picture. Comparing the simulations as well as the effective conductance color maps in Figs.~\ref{fig:BSTS-Kink-90DEG} b) and e) it is clear that the range of angles with reduced conductance becomes narrower for an increased kink angle. This is consistent with the picture based on the Lorentz force inducing a trapping effect via the side facets. The window of angles for which the Lorentz force has opposite sign on the side facets of the two legs of the kink (yielding an opposite trapping effect in each leg and a reduced transmission probability across the kink) naturally narrows down when the legs become more aligned. The relevant angle window corresponds to $\theta_B \in [ \pi/2, \pi] + n\pi$ for the $90^\circ$ kink, and to $\theta_B \in [ 3\pi/2, \pi] + n\pi$ for the $120^\circ$ kink.

\begin{figure}[tb]
\centering
   \includegraphics[width=1\linewidth]{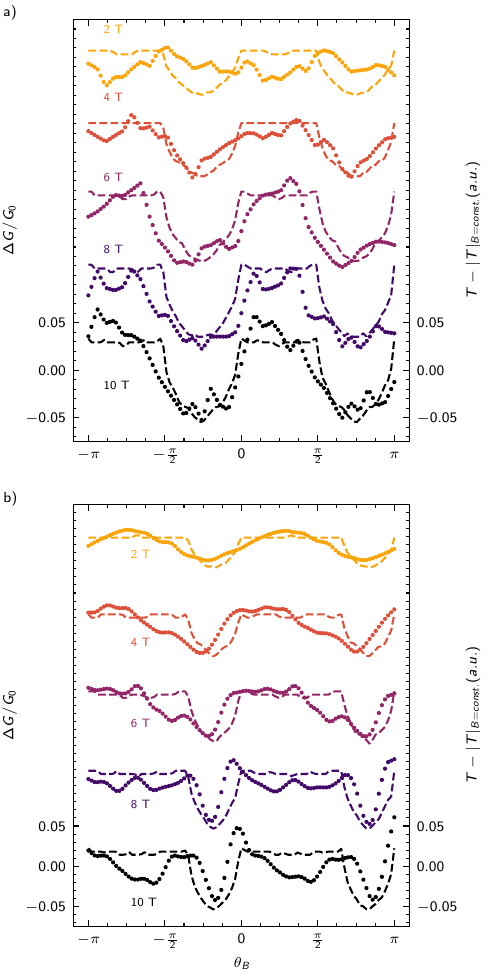}
   \caption{Linecuts of the effective conductance change $\Delta G/G_0$ (dotted lines) and of the normalized transmission $T-|T|_{B=const.}$ (dashed lines) as a function of $\theta_B$ at magnetic fields ranging from 2 to \SI{10}{T} in steps of \SI{2}{T} for the a) \SI{90}{^\circ} and b) \SI{120}{^\circ} kink (taken from Fig.~\ref{fig:BSTS-Kink-90DEG}). Note that the individual curves are offset for better clarity.}
   \label{fig:BSTS-Kink-Linecuts} 
\end{figure}

Compared to our previous study~\cite{Koelzer21}, the high magnetic field ranges of the experiment allow for the sample to get into a regime where the gyroradius is of the order of the ribbon height (approx. $\SI{20}{nm}$). This drastically increases the effective conductance change between the default and suppressed transport regimes to about \SI{8}{\%} (increased by more than an order of magnitude).

In the Supplemental Material of Ref.~\cite{Koelzer21}, it was also shown that the steering effect is only expected for states that are located on the surface, and that the robustness of the effect against disorder is highly enhanced by spin-momentum locking, which suppresses scattering processes with a significant change in momentum. The observed magnetoconductance pattern is therefore not likely to originate from trivial bulk states or trivial surface states without spin-momentum locking. Hence, in-plane magnetic field dependence of the conductance across the kink structures provides a robust transport signature of topological surface states.

\begin{figure}[tb]
    \centering
    \includegraphics[width=\linewidth]{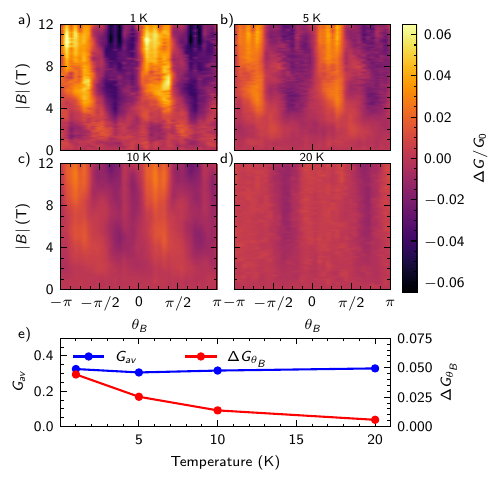}
    \caption{a) to d) Effective normalized conductance $\Delta G/G_0$ of the \SI{90}{^\circ} kink as a function of $\theta_B$ and $B$ at a) 1\,K, b) 5\,K, c) 10\,K, and d) 20\,K. The periodic pattern as a function of in-plane magnetic field angle is clearly vanishing with increasing temperature. e) The peak-to-valley difference $\Delta G_{\theta_B}$ averaged over all magnetic fields in comparison to the average conductance $G_\mathrm{av}$ of the sample averaged over all angles and magnetic fields. These values are displayed as a function of temperature. It is clearly visible that the average conductance of the sample is rather constant with increasing temperature whereas the peak-to-valley difference exponentially drops to zero.} 
    \label{fig:BSTS-Kink-90DEG-Temp}
\end{figure}

For a more detailed comparison of the angle-dependent conductance pattern between the experimental results and the simulations, we take linecuts at certain magnetic fields. Figure~\ref{fig:BSTS-Kink-Linecuts} displays linecuts of the effective conductance change $\Delta G/G_0$ in comparison to linecuts of the normalized transmission at magnetic fields ranging from 2 to \SI{10}{T}, respectively, as a function of in-plane magnetic field angle. Note that the normalized transmission is rescaled and shifted to fit the scale of the conductance changes, in order to compare the angle-dependent $\pi$-periodic pattern. At the linecuts above \SI{2}{T}, simulation and experiment are in very good agreement. For magnetic field strengths of \SI{2}{T} and less, no periodic behaviour of the effective conductance can be seen (cf. Fig.~\ref{fig:BSTS-Kink-90DEG} c)).

Figure~\ref{fig:BSTS-Kink-90DEG-Temp} shows the temperature-dependent effective conductance of the \SI{90}{^\circ} kink for four different temperatures ranging from 1 to 20\,K. The color scheme for all temperatures is normalized to that of the \SI{1}{K} measurement. From previous work performed on similar devices, it was found out that charge carriers are located phase-coherently around the perimeter of a ribbon of similar height up to temperatures of about \SI{20}{K}~\cite{Rosenbach2022}. Therefore, the temperature range allows us to probe the proposed effect in the transition of the phase-coherent regime to a regime where an increased amount of inelastic scattering events occur. The temperature dependence of the effect matches the expected behaviour for an effect that relies on phase coherence around the perimeter of the ribbon.
A vanishing periodic conductance change is visible with the pattern completely disappearing at temperatures above \SI{20}{K}. The impact of the in-plane magnetic field on the conductance due to trapping of surface states on certain side facets of the nanoribbon is only expected to occur when the surface states are phase coherent around the perimeter of the ribbon, i.e., at sufficiently low temperatures. At elevated temperatures, the phase-coherence length reduces so that the surface states on the different side facets are effectively becoming independent and there can be no depletion of surface states on either top or bottom surface of the ribbon through a Lorentz force acting on the side facets.

To emphasize that the trapping effect can be associated with the phase coherence of the surface states, we plot the average conductance $G_\mathrm{av}$ (over all magnetic field strengths and angles) of the device as a function of temperature and compare it with the peak-to-valley difference of conductance over different in-plane magnetic field angles (see Fig. \ref{fig:BSTS-Kink-90DEG-Temp} e)). The peak-to-valley difference $\Delta G_{\theta_B}$ is obtained by taking the difference between the conductance averaged over $\theta_B \in [ 0, \pi/2] + n\pi$ (peak) and over $\theta_B \in [ \pi/2, \pi] + n\pi$ (valley) for all magnetic field strengths $B > 4$ T. While the average conductance of the sample is barely affected by temperature (and phase coherence around the perimeter), the peak-to-valley difference decreases exponentially with temperature towards zero. This is similar to the temperature dependence of Aharonov-Bohm oscillations, for example, $\propto \exp(-P/l_\phi(T))$, with perimeter $P$ and temperature-dependent phase-coherence length $l_\phi(T)$, which also naturally depend on phase coherence around the perimeter~\cite{Behner23}.

Note that an out-of-plane component of the magnetic field, due to a small misalignment between the plane of rotation of the magnetic field and the plane of the kink structure, for example, is unlikely to be responsible for the observed conductance pattern because of the following reasons. First, our setup ensures good alignment between the kink structure and the external magnetic field (a misalignment of at most 2 degrees). Second, such misalignment would yield a sinusoidal dependence on $\theta_B$ for the out-of-plane component, with no qualitative difference for different kink angles (unlike the measured patterns). Furthermore, the observed minima and maxima of the conductance would only match the expected angles (based on the trapping effect) coincidentally. Third, a conventional magnetoresistance related to an out-of-plane magnetic field component would not have a strong temperature dependence that indicates a connection with phase coherence of surface states around the perimeter of the legs of the kink.

\section{Conclusion}

In conclusion, for 3D topological nanoribbon-based $90^\circ$ and $120^\circ$ kink structures we observed pronounced modulations in the conductance upon varying the in-plane magnetic field orientation. From the decrease of the modulation pattern with increasing temperature we deduced that the effect is based on phase-coherent carriers. In a semiclassical picture, the modulations can be explained by an orbital effect trapping electrons in the topological surface states either on the upper or lower surface of the nanoribbon legs, which affects their transmission probability across the kink. Our experimental results are in good agreement with the theoretically expected transport behavior, and consistent with the differences expected for different kink angles. The transport properties of TI nanoribbon-based kinks with an in-plane magnetic field offer interesting perspectives for the design of topological quantum circuits. The orbital effect could be exploited to drive a kink-shaped TI Josephson junction into the topological regime with Majorana states, while also offering a tuning knob for the Josephson energy through modulation of the surface-state transparency.

\section{Acknowledgments}
We thank Herbert Kertz for technical assistance, and Florian Lentz and Stefan Trellenkamp for electron beam lithography. This work was partly funded by the Deutsche Forschungsgemeinschaft (DFG, German Research Foundation) under Germany’s Excellence Strategy - Cluster of Excellence Matter and Light for Quantum Computing (ML4Q) EXC 2004/1 – 390534769. K.M.\ acknowledges the financial support by the Bavarian Ministry of Economic Affairs, Regional Development and Energy within Bavaria’s High-Tech Agenda Project "Bausteine für das Quantencomputing auf Basis topologischer Materialien mit experimentellen und theoretischen Ansätzen" (Grant No.\ 07 02/686 58/1/21 1/22 2/23). P.S.\ and K.M.\ acknowledge financial support by the German Federal Ministry of Education and Research (BMBF) via the Quantum Futur project “MajoranaChips” (Grant No.\ 13N15264) within the funding program Photonic Research Germany.

\clearpage
\widetext

\setcounter{section}{0}
\setcounter{equation}{0}
\setcounter{figure}{0}
\setcounter{table}{0}
\setcounter{page}{1}
\makeatletter
\renewcommand{\thesection}{SUPPLEMENTAL NOTE \arabic{section}}
\renewcommand{\thesubsection}{\Alph{subsection}}
\renewcommand{\theequation}{S\arabic{equation}}
\renewcommand{\thefigure}{S\arabic{figure}}
\renewcommand{\figurename}{Supplemental Figure}
\renewcommand{\bibnumfmt}[1]{[S#1]}
\renewcommand{\citenumfont}[1]{S#1}

\begin{center}
\textbf{Supplemental Material: In-plane magnetic field driven conductance modulations in topological insulator kinks}
\end{center}

\section{EFFECTIVE TRAPPING MODEL}
\begin{figure}[b]
    \centering
    \includegraphics[width=0.99\linewidth]{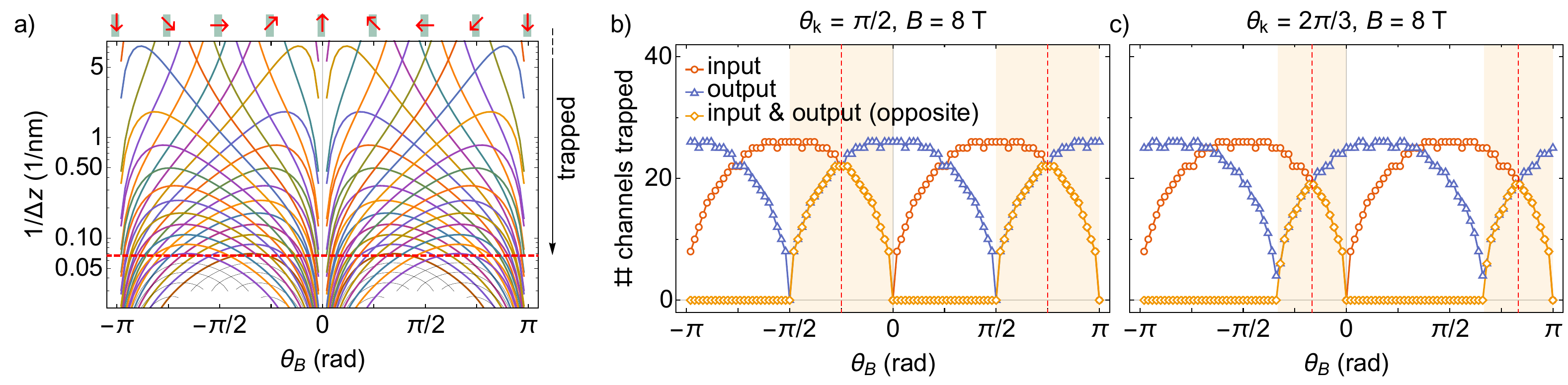}
    \caption{
    a) The transverse extent of the different subbands of a 3D TI nanoribbon as a function of the external magnetic field orientation angle according to Eq.~\eqref{eq:transverse-extent}. The transverse extent of the subbands that become trapped for certain angles ($\Delta z$ smaller than the height of the ribbon, indicated with red dashed line) is indicated in color, and the extent of subbands that do not get trapped for any angle is indicated in black. b), c) The total number of trapped subbands in each leg of a b) 90${}^\circ$ and c) 120${}^\circ$ kink, as well as the number of trapped subbands that is trapped on opposite surfaces in the two legs of the kink. The two intervals of angles in which subbands can get trapped on opposite surfaces are indicated in yellow. Here, we have assumed an (in-plane) rotating external magnetic field of 8\,T.
    }
    \label{fig:S1}
\end{figure}

To explain the origin of the angle-modulated conductance pattern, we consider the impact of the orbital effect on the topological surface states in a semiclassical picture, analogous to the treatment of the steering effect in a topological insulator-based T-junction in Ref.~\cite{Moors18supp}. On the top and bottom facets of the ribbon, the Lorentz force points perpendicular to the surface and can be neglected for surface-state motion that is bound to the surface of the ribbon. On the side facets (assumed to be perpendicular to the plane here), however, the Lorentz force induces a circular motion of the charge carriers with gyro\-radius $R_\mathrm{g} = |E_\textsc{f}/(e B_\perp v_\textsc{d})|$, with $E_\textsc{f}$ the Fermi energy, $B_\perp$ the magnetic field component perpendicular to the side facet, and $v_\textsc{d}$ the Dirac velocity of the topological surface states ($|E_\textsc{f}| = \hbar v_\textsc{d} \sqrt{k_\parallel^2 + k_\perp^2}$). Depending on the transverse velocity of the surface state, the state will be able to traverse the side facet from bottom to top (or vice versa) or not. We can quantify this condition in terms of the transverse extent of the orbital motion $\Delta z$, given by:
\begin{equation} \label{eq:transverse-extent}
    \Delta z = R_\mathrm{g} ( 1 - v_\parallel/v_\perp ) = \left| \frac{E_\textsc{f}}{e|\mathbf{B}| \sin\theta \, v_\textsc{d}} \right| \left( 1 - \frac{|k|}{\sqrt{k^2 + (2 \pi j)^2/P^2}} \right),
\end{equation}
with $v_\parallel$ ($v_\perp$) the velocity component along (perpendicular to) the ribbon, $\theta$ the relative angle between the ribbon and the in-plane magnetic field, $j \equiv l - 1/2 - \Phi/\Phi_0$ the generalized quantum number for the transverse wave number of the topological surface state, $k_\perp = 2 \pi j / P$, including contributions of the transverse orbital motion, the nontrivial Berry phase $\pi$, and the magnetic flux enclosed by the perimeter $P$ of the ribbon cross section in units of the flux quantum $\Phi_0 \equiv h/e$.
When the transverse extent is smaller than the height of the ribbon, the state cannot traverse the side facet (see Fig.~\ref{fig:S1}). For surface states that wrap phase-coherently around the perimeter of the ribbon cross section, it follows that the top or bottom surface effectively gets depleted (or, equivalently, trapped on the opposite surface). When the surface states on the two legs of the kink are trapped on opposite facets of the ribbon, the transmission of these states across the kink is suppressed.

\begin{figure}[tb]
    \centering
    \includegraphics[width=0.9\linewidth]{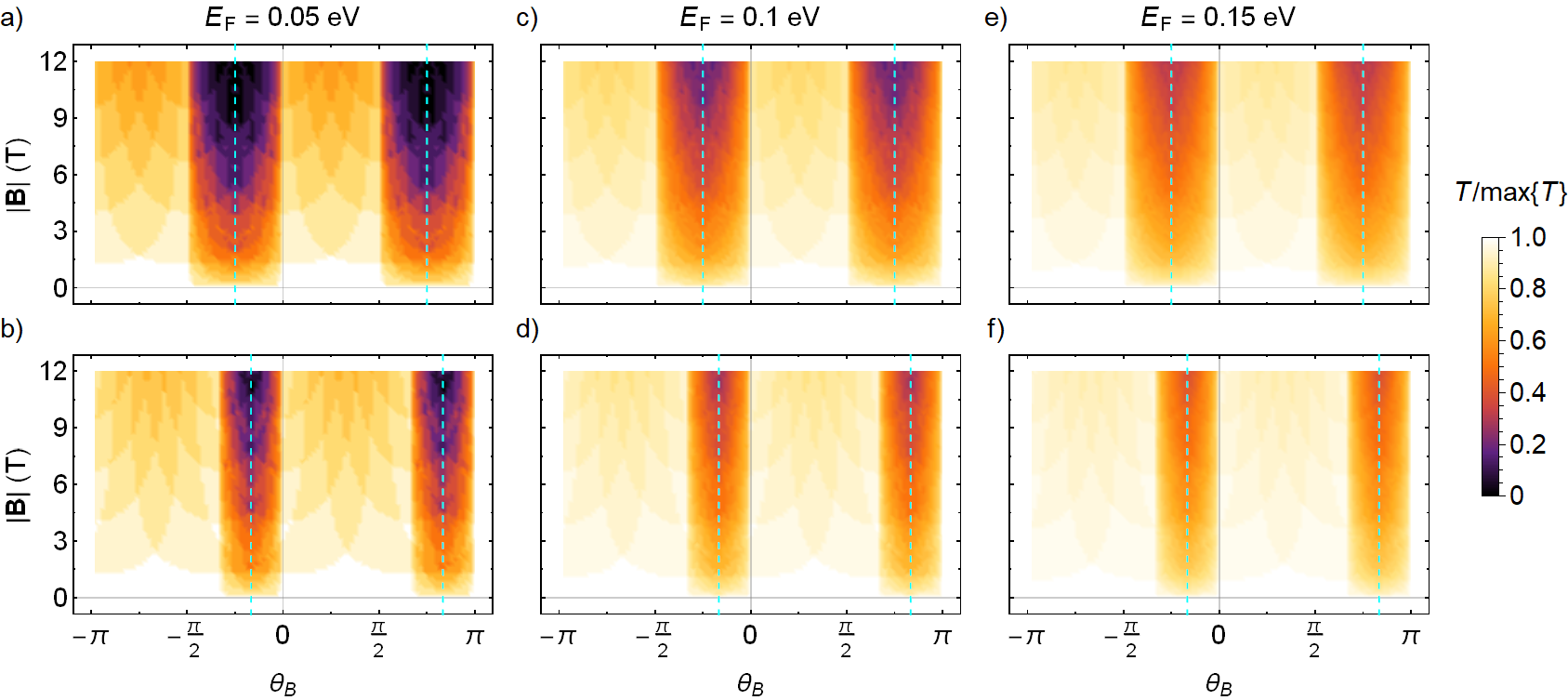}
    \caption{
    a)-f) Normalized transmission as a function of absolute magnetic field $|B|$ and magnetic field angle $\theta_B$ for a)-c) $90^\circ$ and d)-f) $120^\circ$ kink structures and different values of the Fermi energy (relative to the Dirac point). The middle column corresponds to Figs. 2 c) and f) in the Main Text.
    }
    \label{fig:S2}
\end{figure}

For Figs.~2 c) and f) in the Main Text and Fig.~\ref{fig:S1}, we consider a 3D TI nanoribbon with a width of 100 nm and a height of 15 nm, Dirac velocity $v_\textsc{d} = 3.5\times10^5\,$m/s, and Fermi energy $E_\mathrm{F} = 0.1\,$eV (with the Dirac point at $E_\mathrm{F} = 0$ as reference), which are reasonable assumptions for the TI kink structures under consideration in this work. Evaluating the magnetic field-dependent conductance pattern for different values of $E_\mathrm{F}$ (see Fig.~\ref{fig:S2}), we find that the reduction of the conductance is more pronounced near the Dirac point. However, the overall shape of the conductance pattern and, in particular, the range of angles over which the conductance is reduced, is not sensitive to the precise value of the Fermi level. A more detailed comparison between the simulation results and the experimental data is beyond the scope of our model, however, as bulk contributions and disorder may also affect the magnitude of the reduction of the total conductance while being neglected in our model.

\newpage
\section{UNIVERSAL CONDUCTANCE FLUCTUATION MAGNITUDE}
In order to prove that the magnitude of the universal conductance fluctuations (UCFs) \cite{Lee87supp,Koelzer20supp} and their angle dependence have no impact on the appearance of the angle-dependent conductance pattern, we evaluate the fluctuation amplitude of the UCFs as a function of the in-plane magnetic field angle using the $\delta_{G_\mathrm{RMS}}$ value. A smooth background has been subtracted by applying a first-order Savitzky–Golay filter. Hence, only the UCFs are considered in the root-mean-square analysis (RMS) following  $\delta_{G_\mathrm{RMS}} = \sqrt{\sum_{i=1}^{n} x^2_i}$.
\begin{figure}[h]
    \centering
    \includegraphics[width=0.99\linewidth]{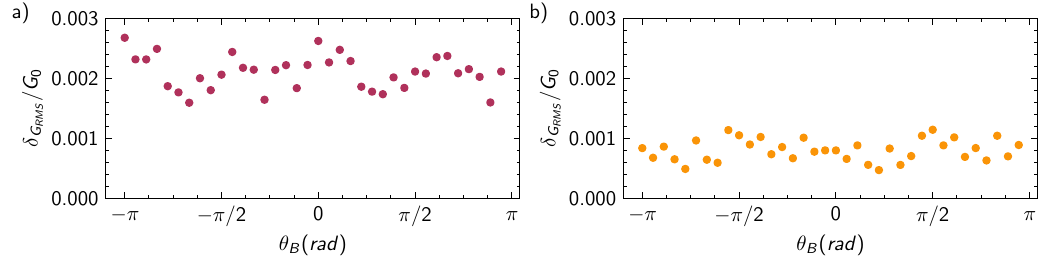}
    \caption{$\delta_{G_\mathrm{RMS}}$ analysis for the measurements as a function of magnetic field angle $\theta_{B}$ performed on the a) $\SI{90}{^\circ}$ and b) $\SI{120}{^\circ}$ kink, respectively. The values are normalized by the conductance quantum $G_0$. WAL leads to a sharp peak around \SI{0}{T}, falsifying the $\delta_{G_\mathrm{RMS}}$ value, if included.  } 
    \label{fig:S3}
\end{figure}
Figure \ref{fig:S3} shows the $\delta_{G_\mathrm{RMS}}$ normalized by the conductance quantum $G_0$ as a function of in-plane magnetic field angle for both the $\SI{90}{^\circ}$ and $\SI{120}{^\circ}$ kink. Note that, data points corresponding to magnetic fields below \SI{0.5}{T} have been excluded from the analysis. Comparing the amplitude of the UCFs with that of the conductance change due to orbital effect-driven trapping, it is clear that the angle-dependent conductance pattern is one order of magnitude larger than that of the UCFs. This further demonstrates that the angle dependent change in device conductance is due to the proposed orbital effect.

\end{document}